\documentclass[aps,pra,reprint,showpacs,footinbib]{revtex4-1}

\usepackage{amsmath}
\usepackage{amsfonts}
\usepackage{amsthm}
\usepackage{graphicx}
\usepackage{enumerate}
\usepackage{color}
\usepackage[T1]{fontenc}
\usepackage{mathptmx}
\usepackage{braket}
\usepackage{todonotes}
\usepackage{soul}
\graphicspath{{figures/}}

\usepackage[bookmarks=false,colorlinks,citecolor=blue,linkcolor=blue,anchorcolor=blue,urlcolor=blue]{hyperref}

\begin{document}

\title{One-sided measurement-device-independent quantum key distribution}

\author{Wen-Fei Cao}
\author{Yi-Zheng Zhen}
\author{Yu-Lin Zheng}
\author{Li Li}
\email{eidos@ustc.edu.cn}
\author{Zeng-Bing Chen}
\email{zbchen@ustc.edu.cn}
\author{Nai-Le Liu}
\email{nlliu@ustc.edu.cn}
\author{Kai Chen}
\email{kaichen@ustc.edu.cn}

\address{Hefei National Laboratory for Physical Sciences at Microscale and Department of Modern Physics, University of Science and Technology of China, Hefei, Anhui 230026, People's Republic of China\\
and CAS Center for Excellence and Synergetic Innovation Center in Quantum Information and Quantum Physics, University of Science and Technology of China, Hefei, Anhui 230026, People's Republic of China}

\date{Jan 16, 2018}

\begin{abstract}
Measurement-device-independent quantum key distribution (MDI-QKD) protocol was proposed to remove all the detector side channel attacks, while its security relies on the trusted encoding systems. Here we propose a one-sided MDI-QKD (1SMDI-QKD) protocol, which enjoys detection loophole-free advantage, and at the same time weakens the state preparation assumption in MDI-QKD. The 1SMDI-QKD can be regarded as a modified MDI-QKD, in which Bob's encoding system is trusted, while Alice's is uncharacterized. For the practical implementation, we also provide a scheme by utilizing coherent light source with an analytical two decoy state estimation method. Simulation with realistic experimental parameters shows that the protocol has a promising performance, and thus can be applied to practical QKD applications.
\end{abstract}


\maketitle

\section{Introduction}\label{sec:introduction}
Quantum key distribution (QKD) \cite{Gisin2002, Scarani2009, Lo2014} enables two distant parties to produce random common secret key bits which can be used for secure communication. In theory, 
the unconditional security of QKD is guaranteed by the laws of quantum mechanics \cite{Lo1999, Shor2000, Renner2005}. In the usual security proof of QKD, it is assumed that all the devices, both sources and measurement devices, are trusted or well-characterized. In a realistic setup, however, the imperfection of the practical devices leads to various kinds of side-channel attacks, including fake-state attack \cite{Makarov2006, Makarov2008}, time-shift attack \cite{Qi2007, Zhao2008}, phase-remapping attack \cite{Fung2007, Xu2010}, detector-blinding attack \cite{Lydersen2010, Gerhardt2011}, dead time attack \cite{Weier2011}, unambiguous state discrimination attack \cite{Tang2013}, etc. \cite{Jain2011, Lydersen2011, Wiechers2011, Jiang2013, Bugge2014, Sun2015}.

As a result, on the route of realizing applications of QKD, it is very crucial to bridge the gap between theoretical expectations and realistic setups. To connect theory with practice, several approaches have been proposed. A once-and-for-all solution is device-independent QKD (DI-QKD) \cite{Mayers1998, Acin2007, Pironio2009, Lim2013, Vazirani2014}, which allows all the devices (i.e., both sources and measurement devices) being untrusted. 
Based on the violation of Bell inequality, DI-QKD can
defend against any possible eavesdropping. 
Although recently loophole-free Bell tests have been carried out \cite{Hensen2015, Giustina2015}, the DI-QKD is still very difficult to be implemented. At the moment, what obstructs its practical application is not only the requirement of high detection efficiency but also the quite low secure key rate at feasible distances.

To make the realistic QKD applications more applicable, one has to make more assumptions on the characterization of devices. For such a purpose, one-sided device-independent QKD (1SDI-QKD) \cite{Branciard2012, Tomamichel2013} has been proposed, where the measurement device on one side is assumed to be trusted. The security of 1SDI-QKD is based on demonstration of EPR steering \cite{Schroedinger1935, Wiseman2007, Jones2007, Reid2009}, in analogy to the demonstration of Bell inequality in DI-QKD. 
The 1SDI-QKD enjoys the advantage of a lower detection efficiency requirement than that of DI-QKD \cite{Branciard2012}, since it is easier to realize a loophole-free EPR steering experiment \cite{Bennet2012, Wittmann2012, Smith2012} than a loophole-free Bell test.
However, the detection efficiency requirement of 1SDI-QKD is still too high for a realistic QKD application, especially when the desired communication distance is long \cite{Branciard2012}. 

Based on the time-reversed entanglement-based QKD \cite{Ekert1991, Biham1996, Inamori2002},
the measurement-device-independent QKD (MDI-QKD) \cite{Lo2012, Braunstein2012}
was proposed to close all kinds of detection side-channel attacks. The MDI-QKD is an attractive scheme for practical implementations because of its high security and long achievable distances. More recently, several MDI-QKD experiments have been successfully carried out \cite{Rubenok2013, Liu2013, FerreiradaSilva2013, Tang2014, Tang2014a, Tang2016, Yin2016}. A crucial assumption of MDI-QKD is that the state preparation device is nearly perfect, while such requirements might not be strictly satisfied in practice. For example, one's encoding device may be manufactured by a untrusted third party.

Furthermore, several protocols have been proposed \cite{Yin2013, Yin2014, Li2014, Zhang2014, Tamaki2014}  to relax the assumptions on the encoding systems. 
By modifying the original MDI-QKD and assuming qubit sources, Yin {\it et al.} \cite{Yin2013, Yin2014} have proved that MDI-QKD can still be secure with uncharacterized encoding systems.
In  \cite{Li2014}, MDI-QKD 
based on the CHSH inequality (CHSH-MDI-QKD)
has been investigated, in which the state is prepared in the two-dimensional Hilbert space.
In \cite{Zhang2014}, the decoy state method was combined with the CHSH-MDI-QKD protocol to guarantee its security when using a weak coherent state source.
In \cite{Tamaki2014}, a general technique that applies to any state preparation flaws in phase-randomized sources was proposed, which has been 
successfully carried out 
in experiments \cite{Xu2015a, Tang2016a}.
With regard to the theoretical analysis of point-to-point quantum communications, the ideal optimal key rates with respect to different scenarios were discussed in Ref. \cite{Pirandola2017}, where all possible local operations and classical communications are taken into account.

In this paper, we propose a one-sided MDI-QKD (1SMDI-QKD) protocol, which enjoys the detection loophole-free advantage, and at the same time weakens the state preparation assumption in MDI-QKD.
The 1SMDI-QKD can be regarded as a modified MDI-QKD, in which Bob's encoding system is trusted, while Alice's encoding system is uncharacterized.
In the single-photon case, Alice's encoding system is assumed to output a quantum state in a two-dimensional Hilbert space basis-independently.
This is the assumption we made about Alice's encoding system besides the common assumptions of the single-photon MDI-QKD case.
For the practical implementation, we also propose a scheme by utilizing coherent light source with an analytical two decoy state estimation method.
Besides, we provide a concise security analysis for both the single-photon case and the decoy-state case, using a virtual-photon qubit idea.
Compared with the existing protocol \cite{Yin2013, Yin2014, Li2014, Zhang2014, Tamaki2014}, the security proof of our protocol is much more straightforward, and the data post-processing is easier to be applied for experimentists.
Simulation with realistic experimental parameters also shows that our protocol has a promising performance, and thus can be applied to practical QKD applications.
Without changing the experiment apparatus, one can achieve a higher security level with fewer assumptions about source device for the existing MDI-QKD experiments \cite{Rubenok2013, Liu2013, FerreiradaSilva2013, Tang2014, Tang2014a, Tang2016, Yin2016}.

The paper is organized as follows. In Sec. \ref{sec:1sdi-qkd-protocol}, we provide a short review on the 1SDI-QKD and a concrete security analysis. In Sec. \ref{sec:single-photon-1smdi-qkd}, we present a single-photon version of 1SMDI-QKD, derive the key rate formula with unconditional security, and make a discussion on the assumptions we used.
In Sec.~\ref{sec:decoy-state-protocol}, we present a decoy-state 1SMDI-QKD. In Sec.~\ref{sec:decoy-state-key-rate}, we derive the secure key rate for decoy-state 1SMDI-QKD. In Sec. \ref{sec:decoy-state-parameter-estimation}, we provide methods for estimating the parameters used in the key rate formula. In Sec. \ref{sec:simulation-results}, we show simulation results for the key rate by comparing between 1SMDI-QKD and MDI-QKD for, both cases of asymptotical and finite decoy states situations. Finally, we give a summary of this paper in Sec. \ref{sec:conclusion}.

\section{One-sided device-independent QKD}\label{sec:1sdi-qkd-protocol}
\subsection{Protocol description} \label{sec:one-sided-protocol}


In this subsection, we introduce the scheme for
the non-post-selected version of 
1SDI-QKD \cite{Branciard2012}.
Following 1SDI-QKD, the non-post-selected version of the 1SDI-QKD protocol can be described as follows.
Alice and Bob receive some quantum systems from an untrusted external source. 
Alice (Bob) can choose from two binary measurement operators, $A_1$ and $A_2$ ($B_1$ and $B_2$).
Alice's measurement device is untrusted, which is treated
as a black box with two possible settings and two possible outputs each time.
Bob's measurement device is trusted, which is assumed to make projective measurements in some qubit subspace.
After performing error correction and privacy amplification, they extract a secret common key string finally.
The difference between the non-post-selected version of the 1SDI-QKD protocol and original 1SDI-QKD is that all the strings of classical bits Alice gets from measurements $A_1$ are used for the key generation without post-selection.

\subsection{Key rate for 1SDI-QKD}\label{sec:one-sided-key-rate}
In this subsection we formulate the key rate for non-post-selected 1SDI-QKD. A post-selected version can also be found in Ref.~\cite{Branciard2012}.
We denote by $\mathbf{A}_i$ and $\mathbf{B}_i$ the strings of classical bits Alice and Bob get from measurements $A_i$ and $B_i$.
From the $N$-bit strings $\mathbf{A}_1$ and $\mathbf{B}_1$, 
Alice and Bob can extract
a secret key of length \cite{Tomamichel2011, Renes2012},
\begin{equation}
l \approx H_{min}^{\epsilon}(\mathbf{B}_1|\mathbf{E})
-H_{max}^{\epsilon}(\mathbf{A}_1|\mathbf{B}_1)
\end{equation}
where $H_{min}^{\epsilon}(\mathbf{B}_1|\mathbf{E})$ denotes the smooth entropy of $\mathbf{B}_1$ conditioned on Eve's information $\mathbf{E}$; $H_{max}^{\epsilon}(\mathbf{A}_1|\mathbf{B}_1)$ denotes the smooth max entropy of $\mathbf{B}_1$ conditioned on $\mathbf{A}_1$.
From the generalized uncertainty relation, one has
\begin{equation}
\begin{split}
H_{min}^{\epsilon}(\mathbf{B}_1|\mathbf{E})
&\geq q N- H_{max}^{\epsilon}(\mathbf{A}_2|\mathbf{B}_2)\\
H_{max}^{\epsilon}(\mathbf{A}_1|\mathbf{B}_1) &\leq N h(e_1)\\
H_{max}^{\epsilon}(\mathbf{A}_2|\mathbf{B}_2) &\leq N h(e_2)
\end{split}
\end{equation}
where $e_i$ is the bit error rate between $\mathbf{A}_i$ and $\mathbf{B}_i$, and $q$ is a parameter to depict how distinct Bob's two measurements are. For orthogonal qubit measurements, $q = 1$.
Finally, one can achieve the key rate $r \doteq l/N$ for non-post-selected 1SDI-QKD as follows
\begin{equation}
r \geq 1- h(e_1) -h(e_2).
\label{eq:key-rate-of-1sdi}
\end{equation}

\section{Single-photon 1SMDI-QKD} \label{sec:single-photon-1smdi-qkd}

In this section, we present a single-photon version of 1SMDI-QKD (see Fig.~\ref{fig:scheme}) in which single-photon sources are used by Alice and Bob. 
We leave a more practical setup using the weak coherent source state for implementation purposes in Sec.~\ref{sec:decoy-state-1smdi-qkd}.

\subsection{Protocol description} \label{sec:single-photon-protocol}
\begin{figure}[hbt]\center
\includegraphics[width=\linewidth]{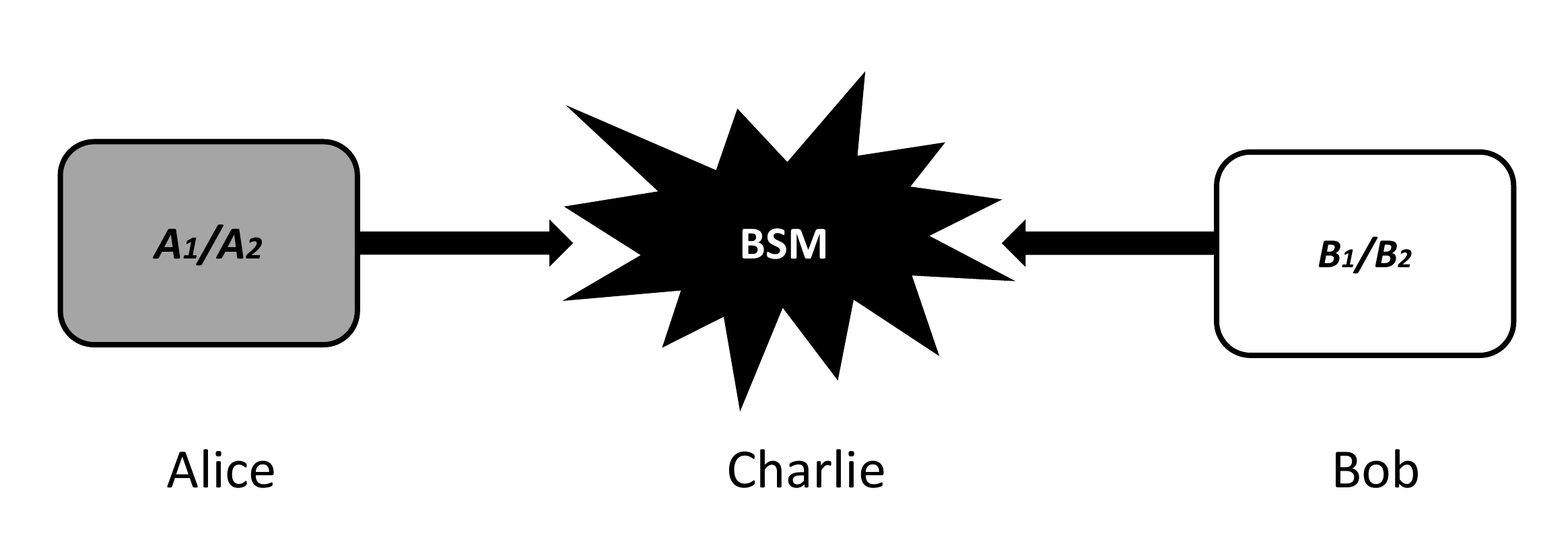}
\caption{\label{fig:scheme}Schematic diagram of 1SMDI-QKD.
BSM denotes Bell state measurement; $A_i$ and $B_i$ ($i \in \{1,2\}$) denote the encoding basis of Alice and Bob.
Alice and Bob prepare some quantum states and send them to an untrusted relay, Charlie, who is supposed to perform BSM and broadcast the results.
Bob's encoding system is trusted, while Alice's is uncharacterized and assumed to output the quantum state in two-dimensional Hilbert space. The white box denotes the trusted device, the black box denotes the untrusted device, and the gray box denotes the uncharacterized device. }
\end{figure}

The single-photon 1SMDI-QKD protocol runs as follows.
Suppose that both Alice and Bob prepare single photons in the four BB84 states ($\ket{0}, \ket{1},\ket{\pm}=1/\sqrt{2}(\ket{0}\pm\ket{1})$, where $\ket{0}, \ket{1}$ are the eigenstates of $\sigma_z$, and $\ket{\pm}$ are the eigenstates of $\sigma_x$) \cite{Bennett1984}.
Alice's encoding system is uncharacterized, i.e., it can be treated as a gray box which receives encoding information and outputs quantum state in two-dimensional Hilbert space (shown in Fig.~\ref{fig:scheme}).
Alice and Bob send photons to the untrusted relay, Charlie, who performs a Bell state measurement (BSM) and projects the received pulses into one of the Bell states ($\ket{\phi^{\pm}}=\frac{1}{\sqrt{2}}\ket{00}\pm\ket{11},\ket{\psi^{\pm}}=\frac{1}{\sqrt{2}}(\ket{01}\pm\ket{10})$). Then Charlie announces the BSM measurement results among a public classical channel.
Afterward, Alice and Bob estimate the quantum phase error and quantum bit error rate (QBER).
They then perform error correction to get a correct key bits string and privacy amplification to remove the information obtained by any possible eavesdropper Eve.
Finally, they obtain a correct and secure key bits string that can be used for later secure communication.

\subsection{Key rate for single-photon 1SMDI-QKD}\label{sec:single-photon-key-rate}

\begin{figure}[hbt]\center
\includegraphics[width=\linewidth]{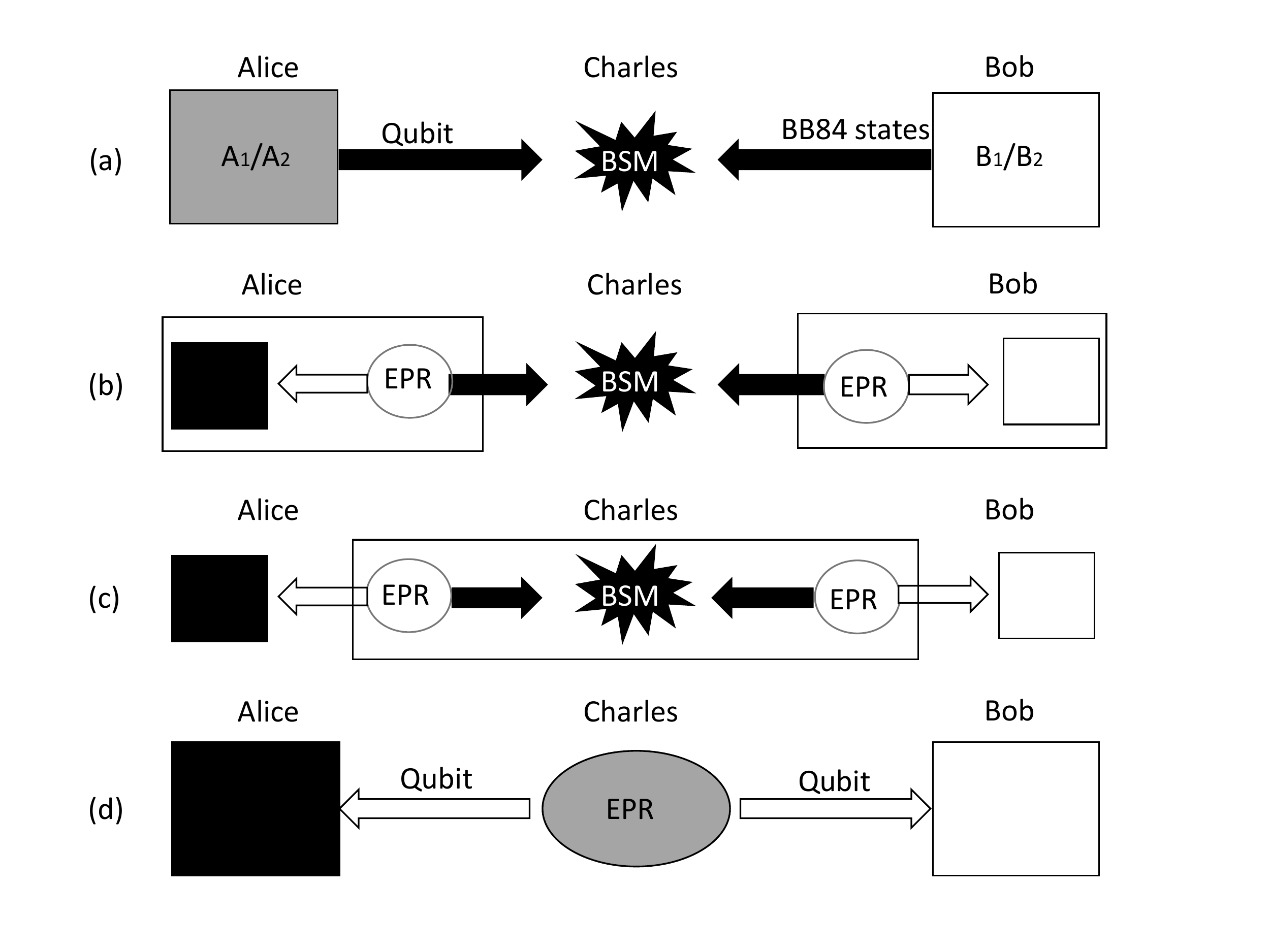}
\caption{\label{fig:security-analysis}
EPR denotes Einstein-Podolsky-Rosen state; BSM denotes the Bell state measurement.
(a) Single-photon 1SMDI-QKD in which Alice's encoding system is uncharacterized.
(b) Virtual-photon single-photon 1SMDI-QKD using the EPR source. 
Instead of preparing a BB84 state, Alice and Bob prepares an perfect EPR pair and sends one particle to an untrusted relay, Charlie, who performs BSM and broadcasts the measurement results. 
Alice and Bob measure their particles using the $A_1/A_2$ and $B_1/B_2$ bases, respectively.
Alice's EPR source is perfect while the local detector is untrusted. This is equivalent to saying that Alice's encoding system is uncharacterized, implying the qubit state from
Alice's encoding system [see Fig.~2~(a)] is basis independent.
(c) Time-reversed single-photon 1SMDI-QKD.
Since their measurement operations are commutable, the order of the measurements can be reversed. That is, Alice and Bob can perform the projective measurement after the BSM is performed.
(d) Equivalent EPR-based 1SDI-QKD protocol.
This is exactly the 1SDI-QKD in which an EPR source outputs qubits to Alice and Bob, and its security has been proven in Sec.~\ref{sec:1sdi-qkd-protocol}.
The black, white, and gray boxes denote the untrusted, trusted, and uncharacterized parts, respectively. }
\end{figure}

Similar to MDI-QKD, here we consider a virtual-photon qubit idea (see Fig.~\ref{fig:security-analysis}). 
Alice's encoding system can be treated as a trusted EPR source with an untrusted projective measurement. Bob's encoding system can be treated as a trusted EPR source with a trusted projective measurement. Using the time-reversed idea, we can perform the projective measurement on Alice's and Bob's side after the BSM is performed. Thus, we can treat the two EPR sources and the BSM in the middle as an untrusted EPR source which outputs qubits to Alice and Bob. This is exactly the
1SDI-QKD in which an EPR source outputs qubits to Alice and Bob.
Therefore one can get the key rate for our 1SMDI-QKD directly from Eq.~(\ref{eq:key-rate-of-1sdi}) for the non-post-selected 1SDI-QKD, which reads
\begin{equation}
r \geq 1- h(e_1) -h(e_2)
\end{equation}
where $e_1$ and $e_2$ is the bit error rate in the $\mathbf{Z}$ and $\mathbf{X}$ bases, respectively.

\subsection{Discussions on the assumptions}
\label{sec:single-photon-assumption}
An important assumption in our single-photon 1SMDI-QKD is that the dimension of Alice's encoding system output is fixed.
By taking an example of dimension two (i.e. qubit), we will prove that this assumption is necessary for security purposes.
We will demonstrate that when the dimension of Alice's encoding system output is not fixed, the protocol will be totally insecure. We prove this by constructing a specific attack scheme, in which Charlie can get all the information without introducing any error. Indeed, if the dimension of  Alice's encoding system is not fixed,
it can encode the four BB84 states into four orthogonal states in four-dimensional Hilbert space (for instance, four Bell states). So Charlie can simply perform a four-dimensional projective measurement to distinguish the four orthogonal states.
Then Charlie can perform a projective measurement on the photon sent by Bob according to the states sent by Alice. To be specific, when Alice sends state $\ket{0}$, Charlie performs the $\mathbf{Z}$ basis projective measurement, and the measurement result is denoted by $M_z$. If $M_z=+$, then Charlie reports $\ket{\phi^+}$, $\ket{\phi^-}$ randomly with equal probability; if $M_z=-$, then Charlie reports $\ket{\psi^+}$, $\ket{\psi^-}$ randomly with equal probability.
From Table \ref{tab:4-dimensional-attack} we can see that Charlie can simulate the same probability tables that can be obtained when genuine BB84 states sent by Alice and actual BSM performed by Charlie.
Therefore, Alice and Bob cannot distinguish whether Alice's encoding system sends genuine BB84 states or four orthogonal four-dimensional states. 
That is to say, Charlie can get all the information without introducing any error.
Thus the security can't be guaranteed. To remove this kind of attack, one must restrict that the dimension of Alice's encoding system output is fixed. 
The qubit assumption is commonly made in various QKD protocols such as decoy-state BB84 \cite{Hwang2003, Lo2005, Wang2005} and MDI-QKD \cite{Lo2012}. 
In fact an experimental method for verifying the qubit assumption can be easily implemented as proposed in Ref.~\cite{Xu2015a}. 
To guarantee the qubit assumption, ones needs only to verify that Alice's encoding system has the same mode except in the encoding degree of freedom.
For example, Alice's phase modulator should have the same timing, spectral, spatial, and polarization mode for different encoding phases in the phase-encoding system.

An uncharacterized encoding system can be treated as a perfect EPR source with an untrusted measurement, which implies that the qubit state from Alice's encoding system is basis independent.
To check the basis-independent assumption in experiment, one needs to test the fidelity between the states sent out by Alice $\rho_A^Z$ and $\rho_A^X$, where either the $X$ or $Z$ basis is used. 
If the fidelity is close to unity, one can accept the basis-independent assumption, and vice versa.
Considering that most errors come from the inaccuracy or uncharacterized polarization modulator in the realistic experiment, the basis independent is satisfied in most cases.

\begin{table}[htb]\center
\caption{\label{tab:4-dimensional-attack}
List of possible clicks for the case where Alice sends $\ket{+z}$ and Bob sends one of the four BB84 states. No loss is considered.}
\vspace{3mm}
\setlength{\tabcolsep}{3.5mm}
\begin{tabular}{cccccc}
\hline\hline
Alice & Bob & Possible clicks & $M_z=+$ & $M_z=-$ \\
\hline
$\ket{0}$ & $\ket{0}$ & $\ket{\phi^+}$, $\ket{\phi^-}$ & 1 & 0 \\
$\ket{0}$ & $\ket{1}$ & $\ket{\psi^+}$, $\ket{\psi^-}$ & 0 & 1 \\
$\ket{0}$ & $\ket{+}$ & All & 1/2 & 1/2 \\
$\ket{0}$ & $\ket{-}$ & All & 1/2 & 1/2 \\
\hline\hline
\end{tabular}
\end{table}

\section{Decoy-state 1SMDI-QKD} \label{sec:decoy-state-1smdi-qkd}

Since the single-photon source is difficult to realize in a real experiment mainly due to high cost, the weak coherent state is widely used in quantum information processing tasks. 
However, the nonzero probability of multiphoton pulses in the weak coherent pulses may cause the photon-number-splitting (PNS) attack \cite{Huttner1995, Brassard2000}.
Hence, the decoy-state method \cite{Hwang2003, Lo2005, Wang2005} is employed 
to defeat the multiphoton events in the weak coherent states.
In the decoy-state 1SMDI-QKD, one naturally assumes that
the single-photon portion of the coherent pulses is basis-independent qubit.
In this section, we will present the decoy-state 1SMDI-QKD using the coherent light as the source, derive the secure key rate for decoy-state 1SMDI-QKD, and provide practical methods for estimating the parameters.

\subsection{Protocol dsescription} \label{sec:decoy-state-protocol}

\begin{figure}[hbt]\center
\includegraphics[width=\linewidth]{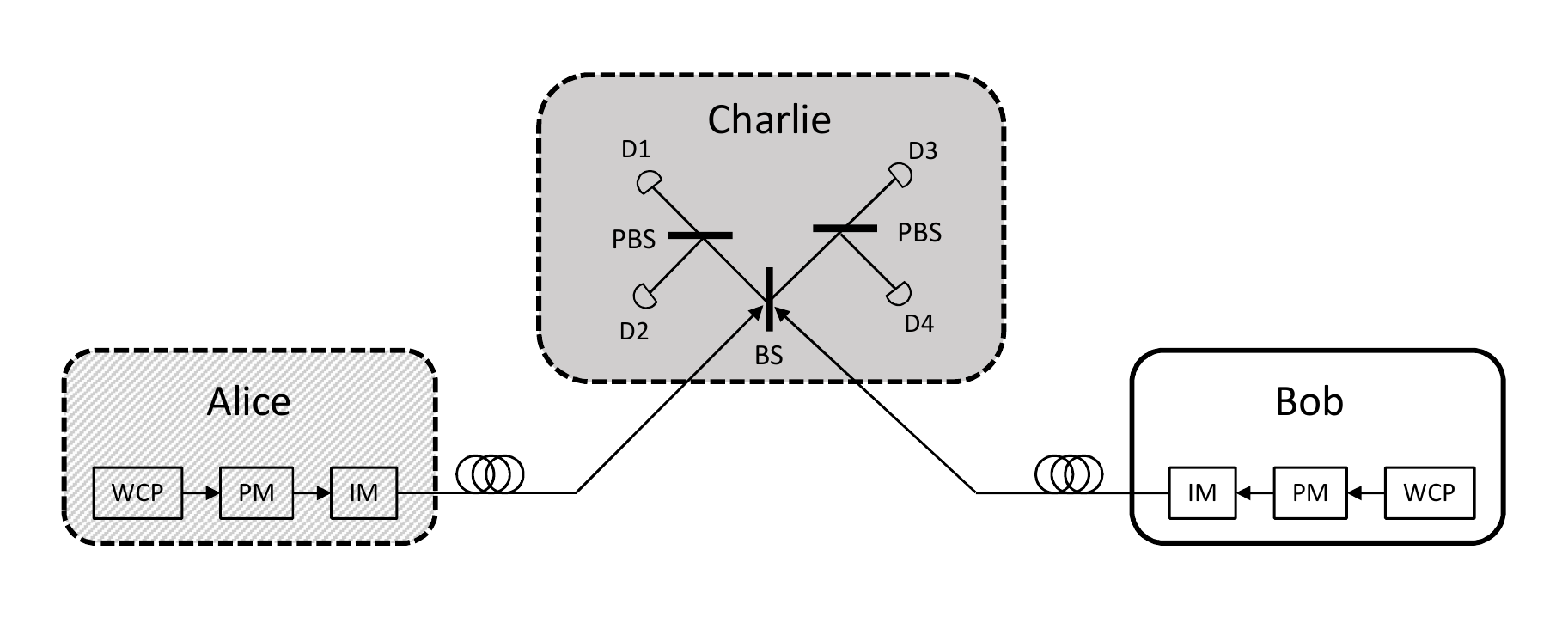}
\caption{\label{fig:decoy-scheme}Basic setup of a decoy-state 1SMDI-QKD protocol.
WCP denotes the weak coherent pulse, PM denotes the polarization modulator, IM denotes the intensity modulator, BS denotes beam splitter, PBS denotes polarization beam splitter, and D1, D2, D3, and D4 denote single-photon detectors.
Alice and Bob prepare WCPs in a different BB84 polarization state randomly, then send them to Charlie to perform the Bell state measurement.
A click in $D1$ and $D4$, or in $D2$ and $D3$, indicates a projection into the Bell state $\ket{\psi^{-}}=(\ket{HV}-\ket{VH})/\sqrt{2}$, and a click in $D1$ and $D2$, or in $D3$ and $D4$ indicates a projection into the Bell state $\ket{\psi^{+}}=(\ket{HV}+\ket{VH})/\sqrt{2}$.
The black, white, and dashed boxes denote the untrusted, trusted, and uncharacterized parts, respectively. 
}
\end{figure}

The basic setup of the decoy-state 1SMDI-QKD using polarization encoding is illustrated in Fig. \ref{fig:decoy-scheme}. The decoy-state 1SMDI-QKD protocol runs as follows.
Suppose that both Alice and Bob prepare random BB84 states with phase randomized weak coherent pulses (WCPs)
, in combination with decoy states.
They send the pulses to the untrusted relay, Charlie, who performs the Bell state measurement (BSM) that projects the received states into a Bell state. Then Charlie announces his BSM measurement results among a public classical channel.
Afterward, Alice and Bob estimate the gain and QBER of single-photon contributions using the decoy state method \cite{Lo2005, Wang2005}.
Finally, they generate a correct and secure key bits string after performing error correction and privacy amplification.

%


\subsection{Key rate for decoy-state 1SMDI-QKD}\label{sec:decoy-state-key-rate}

In this subsection, we present a concise security analysis and derive the key rate for decoy-state 1SMDI-QKD.
Since we have proven the security of single-photon 1SMDI-QKD, here we extend it to the coherent light source situation following the idea of GLLP methods \cite{Gottesman2004}.
From the information theory, the key rate is lower bounded by \cite{Renner2005}
\begin{equation}
R = I(A:B)-\chi(B:E),
\end{equation}
where $I(A:B)$ denotes the mutual information between Alice and Bob, and $\chi(B:E)$ denotes the possible information of Eve.
The first term $I(A:B)$
quantifies the amount of classical information between Alice and Bob after error correction.
The second term $\chi(B:E)$ estimates Eve's knowledge on the raw key, which will be reduced to an arbitrarily small amount
after privacy amplification.

Denote by $\mathbf{A}_i$ and $\mathbf{B}_i$ the random bits of Alice and Bob post-selected based on a successful Bell measurement along $A_i$ and $B_i$ bases for $i=1,2$. The $A_1$, $B_1$ denotes the $\mathbf{Z}$ basis, $A_2$, $B_2$ denotes the $\mathbf{X}$ basis when BB84 states are used. Here we assume that the final key is extracted from the data measured in the $\mathbf{Z}$ basis. 
The mutual information between $\mathbf{A}_1$ and $\mathbf{B}_1$, considering the information leaked in the error-correction process is given by
\begin{equation}
I(\mathbf{A_1}:\mathbf{B_1}) =  1 - f \cdot H(E_{\mu\nu}^{ZZ}),
\end{equation}
where $E_{\mu\nu}^{ZZ}$ denotes the QBER with intensities $\mu$ and $\nu$ when Alice and Bob both use the $\mathbf{Z}$ basis, $\mu$($\nu$) is the mean photon number of Alice's (Bob's) signal state, and $f>1$ is the error correction inefficiency for the error correction process. Denote by $Q_{\mu\nu}^{ZZ}$ the overall gain with intensities $\mu$ and $\nu$ when Alice and Bob both use the $\mathbf{Z}$ basis, 
then the mutual information between Alice and Bob 
is given by 
\begin{equation}
I(A:B) = Q_{\mu\nu}^{ZZ} \cdot  (1 - f \cdot H(E_{\mu\nu}^{ZZ})),
\end{equation}

Following the idea of GLLP methods \cite{Gottesman2004}, all the possible information of Eve can be divided into \textit{tagged} and \textit{untagged} portions,
in which the tagged portion comes from the multiphoton pulses, while the untagged portion comes from the single-photon pulses.
So the information of Eve 
can be written as
\begin{equation}
\begin{split}
\chi(B:E) &= \chi^{t}(B:E) + \chi^{u}(B:E),\\
\chi^{t}(B:E) &= Q_{\mu\nu}^{ZZ}-Q_{11}^{ZZ},\\
\chi^{u}(B:E) &= Q_{11}^{ZZ} \cdot H(e_{11}^{XX}),
\end{split}
\end{equation}
where the superscripts \textit{t} and \textit{u} denote tagged and untagged portions, respectively. $Q_{11}^{ZZ}$ denotes the overall gain in the $\mathbf{Z}$ basis, and $e_{11}^{XX}$ denotes the bit error rate of the $\mathbf{X}$ basis when Alice and Bob sends a single photon.

Finally, we derive the lower bound of the secure key rate as follows
\begin{equation}\label{eq:key-rate}
R=Q_{11}^{ZZ}(1 - H(e_{11}^{XX}))-Q_{\mu\nu}^{ZZ}\cdot f \cdot H(E_{\mu\nu}^{ZZ}).
\end{equation}
In a realistic experiment, $Q_{\mu\nu}^{ZZ}$ and $E_{\mu\nu}^{ZZ}$ can be directly
obtained from the experimental measurements results, while
$Q_{11}^{ZZ}$ and $e_{11}^{XX}$ can be estimated by the decoy method, which will be illustrated in the next subsection.

\subsection{Parameter estimation}\label{sec:decoy-state-parameter-estimation}

For simulation purposes, we evaluate the overall gain and QBER when Alice and Bob prepare phase-randomized WCPs. The overall gain and QBER, 
in the situation without eavesdropping, are the same as MDI-QKD, and can be written as follows \cite{Ma2012}:
\begin{equation}
\begin{aligned}
Q_{\mu\nu}^{XX}&=2y^2(1+2y^2-4y I_0(x)+I_0(2x)),\\
E_{\mu\nu}^{XX}Q_{\mu\nu}^{XX}&=e_0 Q_{\mu\nu}-2(e_0-e_d)y^2(I_0(2x)-1),
\end{aligned}
\end{equation}
and
\begin{equation}
Q_{\mu\nu}^{ZZ} =Q_C^Z+Q_E^Z, \quad
E_{\mu\nu}^{ZZ}Q_{\mu\nu}^{ZZ} =e_d Q_C^Z+(1-e_d)Q_E^Z,
\end{equation}
where
\begin{equation}
\begin{aligned}
Q_C^Z =& 2(1-p_d)^2 e^{-\frac{\omega}{2}}(1-(1-p_{d})e^{-\frac{\mu\eta_{a}}{2}})\times(1-(1-p_d)e^{-\frac{\nu\eta_b}{2}}),\\
Q_E^Z =& 2p_d(1-p_d)^2 e^{-\frac{\omega}{2}}(I_0(2x)-(1-p_d)e^{-\frac{\omega}{2}}).\\
\end{aligned}
\end{equation}

In the above equations, $Q_C^Z$, $Q_E^Z$ denote the gains from the correct and false BSM results, respectively.  $I_0(\cdot)$ is the first kind modified Bessel function, $e_d$ represents the misalignment error probability,  $p_d$ is the background count rate, $e_0=1/2$, $\omega=\mu\eta_{a}+\nu\eta_{b}$, $x=\frac{\sqrt{\mu\nu\eta_{a}\eta_{b}}}{2}$, $y=(1-pd)e^{-\omega/4}$, and $\eta_{a}=\eta_{b}=\eta_{d}\times10^{-\alpha L/20}$ is the total efficiency (both channel transmittance efficiency and detection efficiency $\eta_{d}$ included) for Alice and Bob, respectively.

Without Eve's intervention, the yield in $\mathbf{Z}$ basis $Y_{11}^{ZZ}$ and bit error rate with single-photon states in $\mathbf{X}$ basis $e_{11}^{XX}$ are given as fllows:
\begin{equation}
\begin{aligned}
Q_{11}^{ZZ}=& \mu\nu e^{-\mu-\nu}Y_{11}^{ZZ},\\
Y_{11}^{ZZ}=& Y_{11}^{XX}= (1-p_{d})^2(\frac{\eta_{a}\eta_{b}}{2}+(2\eta_{a}+2\eta_{b}-3\eta_{a}\eta_{b})p_{d}\\
&+4(1-\eta_{a})(1-\eta_{b})p_{d}^2),\\
e_{11}^{XX}Y_{11}^{XX}=& e_{0}Y_{11}^{XX}-(e_{0}-e_{d})(1-p_{d})^2\frac{\eta_{a}\eta_{b}}{2}.
\end{aligned}
\end{equation}


In a practical experiment, the length of the raw key and the number of decoy states are finite.
Here, we consider 
a vacuum + weak-decoy-state method
to obtain $Q_{11}^{ZZ}$ and $e_{11}^{XX}$.
In general, Alice and Bob can use $\mu_0$, $\nu_0$, $\mu_1$, $\nu_1$ as the decoy state, and $\mu_2$, $\nu_2$ as the signal state, in which $\mu_{2}=\nu_{2}>\mu_{1}=\nu_{1}>\mu_{0}=\nu_{0}=0$.
The lower bound of $Y_{11}^{ZZ}, Y_{11}^{XX}$ and the upper bound of $e_{11}^{XX}$ are given as follows \cite{Xu2013} :
\begin{small}
\begin{align}
Y_{11}^{L} \geq&\frac{1}{\mu_{2}^2\mu_{1}^2(\mu_{2}-\mu_{1})}(\mu_{2}^3(e^{2\mu_{1}}Q_{\mu_{1}\mu_{1}}^{}+Q_{00}^{}-e^{\mu_{1}}Q_{\mu_{1}0}^{}-e^{\mu_{1}}Q_{0\mu_{1}}^{})\nonumber\\
&-\mu_{1}^3(e^{2\mu_{2}}Q_{\mu_{2}\mu_{2}}^{}+Q_{00}^{}-e^{\mu_{2}}Q_{\mu_{2}0}^{}-e^{\mu_{2}}Q_{0\mu_{2}}^{})), \label{eq:two-decoy-Y11}\\
e_{11}^{XXU} \leq&\frac{1}{\mu_1\nu_1Y_{11}^{XX}}(Q_{00}^{XX}E_{00}^{XX}+e^{\mu_1+\nu_1}Q_{\mu_1\nu_1}^{XX}E_{\mu\nu}^{XX}\nonumber\\
&-e^{\mu_1}Q_{\mu_10}^{XX}E_{\mu_10}^{XX}-e^{\nu_1}Q_{0\nu_1}^{XX}E_{0\nu_1}^{XX}),
\end{align}
\end{small}
in which Eq.~(\ref{eq:two-decoy-Y11}) is applicable to the $\mathbf{Z}$ and $\mathbf{X}$ bases.

We add a trustworthiness parameter $\eta_s$ to depict the trustworthiness of Alice's source abstractly.
This parameter results from the imperfect preparation of states on Alice's side.
The $\eta_s$ denotes the probability of Alice's encoding system outputting the exact BB84 states.
For the case Alice doesn't send the right state, we assume that $50\%$ error is introduced in terms of the worst case. We note that the assumption of $50\%$ error is equivalent to the assumption of white noise on Alice's encoding device. Considering the trustworthiness of Alice's source, 
the gains and QBERs used for simulation can be written as
\begin{equation}
\begin{aligned}
Q_{\mu\nu}^{ZZ'} &= Q_{\mu\nu}^{ZZ}, \quad Q_{11}^{ZZ'}  = Q_{11}^{ZZ} \\
E_{\mu\nu}^{ZZ'} &= \eta_s E_{\mu\nu}^{ZZ} +\frac{1}{2}(1-\eta_s)\\
e_{11}^{XX'}  &= \eta_s e_{11}^{XX} +\frac{1}{2}(1-\eta_s).
\end{aligned}
\end{equation}

\section{Simulation results}\label{sec:simulation-results}
We consider a setup of our proposal with practical experimental parameters from Ref.~\cite{Tang2014}, which are listed in Table \ref{tab:parameters}. We assume that all detectors have the same dark count rates and the same detection efficiencies.
For simplicity, we only consider the asymptotic data case.
One can extend the analysis to the finite-data case by following the procedures in Refs.~\cite{Curty2014, Xu2014}.

\begin{table}[!hbt]\centering
\caption{\label{tab:parameters}
List of experimental parameters used for simulation.
$\eta_{d}$ is the detection efficiency;  $e_{d}$ is the misalignment-error probability of the system; $p_{d}$ is the dark count rate of the detector; $f$ is error correction efficiency; $\alpha$ is the intrinsic loss coefficient of the standard telecom fiber channel.}
\vspace{3mm}
\setlength{\tabcolsep}{4.1mm}
\begin{tabular}{ccccc}
\hline\hline
$\eta_d$ & $e_d$ & $p_d$ & $f$ & $\alpha(dB/km)$\\
\hline
$40\%$ & $1.5\%$ & $3\times10^{-6}$ & $1.16$ & $0.2$\\
\hline\hline
\end{tabular}
\end{table}

\begin{figure}[!hbt]\center
\includegraphics[width=\linewidth]{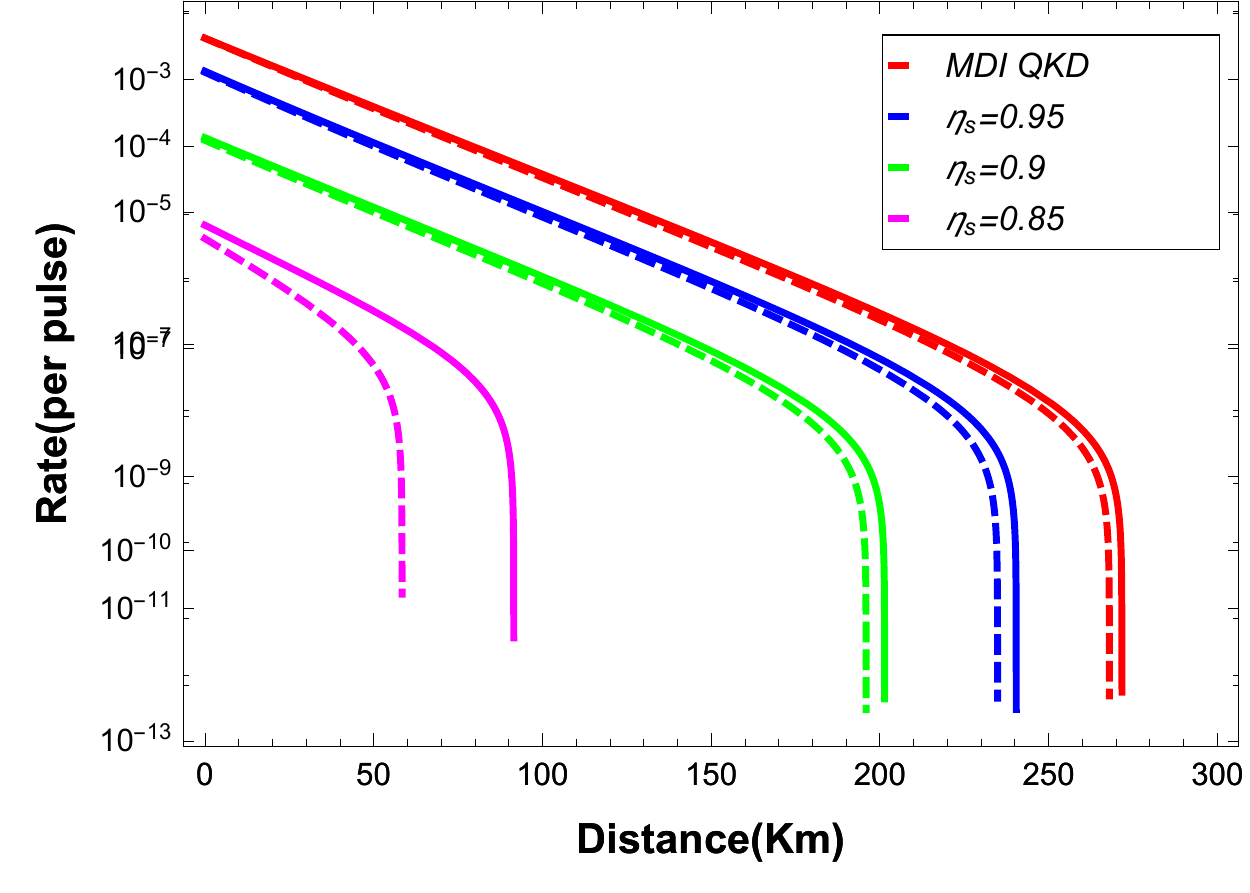}
\caption{\label{fig:key-rate-vs-distance} (Color online) Lower bound on the secret key rate $R$ versus communication distance between Alice and Bob.
The experimental parameters used are listed in Table \ref{tab:parameters}.
The solid line denotes the asymptotic case, and the dashed line denotes the two decoy states case.
The different color corresponds to different $\eta_s$.
In particular, the red line denotes $\eta_s=1$, i.e., the MDI-QKD protocol. The intensities of the signal state used by Alice and Bob are optimized for the asymptotic case, and are $0.45, 0.3, 0.1, 0.05$, for $\eta_s=1, 0.95, 0.9, 0.85$, respectively. The intensity of the decoy states are fixed at $0.01$ and $0$.}
\end{figure}

The secure key rates of the 1SMDI-QKD in the asymptotic case and two decoy states case with different source efficiencies $\eta_s$ are shown in Fig. \ref{fig:key-rate-vs-distance}.
The simulation results of the MDI-QKD protocol are also illustrated with the red curve in the figure for comparison.
We can find that the secure key rate and largest secure communication length of 1SMDI-QKD are slightly lower than that of MDI-QKD, both in the asymptotical case and finite decoy states case.
Moreover, the secure key rate and the largest secure communication distance decrease when the source efficiency $\eta_s$ decreases.
This makes sense since 1SMDI-QKD requires fewer security assumptions than MDI-QKD. The lower largest secure communication distance is the price to pay in order to make Alice's device more uncharacterized.
The simulation results show that 1SMDI-QKD can tolerate high-loss and low trustworthiness of Alice's encoding system.


\section{Conclusion}\label{sec:conclusion}
In summary, we have provided a 1SMDI protocol, which enjoys the detection loophole-free advantage, and at the same time weakens the state preparation assumption in MDI-QKD.
For the practical implementation, we also provide a scheme by utilizing coherent light source with an analytical decoy state method.
The simulation results show that our protocol has a promising performance, and thus can be applied to real-life QKD applications.
Besides, our proposal can be implemented with standard linear optical elements with low detection efficiency over a high-loss channel.
Therefore, it is unnecessary to modify the existing MDI-QKD experiment apparatus, except to guarantee that Alice's encoding states are in two-dimensional Hilbert space.
With the merit of lower requirement, we believe that our proposal is a significant improvement for MDI-QKD under the more realistic situation and paves the way towards the implementations of fully DI-QKD.

\section*{Acknowledgments}

We thank Valerio Scarani for valuable discussion, in particular for bringing to our attention and valuable help for the matters of Sec. \ref{sec:single-photon-assumption}. We also thank Howard M. Wiseman, Feihu Xu, Xiong-Feng Ma, Xiang-Bin Wang, Fei Gao, Tian-Yin Wang, Qiong-Yi He, Jing-Ling Chen, Si-Xia Yu, Zhen-Sheng Yuan, Sheng-Jun Wu, Qiang Zhang, Yu-Ao Chen, Teng-Yun Chen, Jun Zhang and Xiao-Hui Bao for very valuable and enlightening discussions. This work has been supported by the Chinese Academy of Science, the National Fundamental Research Program, and the National Natural Science Foundation of China (Grants No. 11575174, No. 11374287, No. 61125502, and No. 11574297).

\bibliographystyle{apsrev4-1}
\bibliography{bib/1smdi-qkd}

\end{document}